\documentclass[prd,nofootinbib,reprint,twocolumn]{revtex4-1}
\usepackage{amsmath,amssymb,bm,epsfig,color,graphicx}
\usepackage[mathscr]{eucal}
\usepackage{slashed}
\usepackage{cancel}
\usepackage[colorlinks=true,linkcolor=blue,citecolor=blue,urlcolor=blue]{hyperref}
\usepackage[caption=false]{subfig}
\usepackage{hyperref}
\usepackage{tikz,xcolor}
\usepackage{amsmath}
\usepackage{slashed}
\usepackage{comment}
\usepackage{cancel}
\usepackage{multirow}
\usepackage{slashed}

\renewcommand{\thefootnote}{\fnsymbol{footnote}}

\newcommand{\bea}{\begin{array}}
\newcommand{\eea}{\end{array}}
\newcommand{\beq}{\begin{eqnarray}}
\newcommand{\eeq}{\end{eqnarray}}

\definecolor{orange}{RGB}{255,100,0}
\definecolor{rosepink}{RGB}{248,100,100}

\begin{document}
\rightline{UT-Komaba/23-6}

\vspace{-0.5cm}

\title{
\vspace{0.5cm}
NANOGrav Signal from a Dark Conformal Phase Transition
}

\author{
Kohei Fujikura$^{1}$,\footnote{
E-mail address: kfujikura@g.ecc.u-tokyo.ac.jp} 
Sudhakantha Girmohanta$^{2,3}$,\footnote{
E-mail address: } 
Yuichiro Nakai$^{2,3}$\footnote{
E-mail address: ynakai@sjtu.edu.cn} 
and Motoo Suzuki$^{4,5}\footnote{
E-mail address: }$\\*[10pt]
$^1${\it \normalsize
Graduate School of Arts and Sciences, University of Tokyo, Komaba,\\ Meguro-ku, Tokyo 153-8902,
Japan} \\*[3pt]
$^2${\it \normalsize 
Tsung-Dao Lee Institute, Shanghai Jiao Tong University, \\ 520 Shengrong Road, Shanghai 201210, China} \\*[3pt]
$^3${\it \normalsize 
School of Physics and Astronomy, Shanghai Jiao Tong University, \\ 800 Dongchuan Road, Shanghai 200240, China} \\*[3pt]
$^4${\it \normalsize 
Department of Physics, Harvard University, Cambridge, MA 02138, USA} \\*[3pt]
$^5${\it \normalsize
Institute of Particle and Nuclear Studies,\\
	        High Energy Accelerator Research Organization (KEK), Tsukuba 305-0801, Japan}  \\*[5pt]
}

\begin{abstract}

We explore the possibility that a confining first-order phase transition of a nearly-conformal dark sector generates 
the reported NANOGrav signal of a stochastic gravitational wave background.
The visible Standard Model (SM) sector and the dark sector are initially thermally decoupled so that
their temperatures are different.
The nearly conformal phase transition is described by the shallow potential of a dilaton
(or a radion in the 5D holographic perspective)
generated by a new dark Yang-Mills field coupled to the conformal sector.
For a dark sector only gravitationally connected with the visible sector,
the NANOGrav signal is explained by the phase transition without contradicting the $\Delta N_{\rm eff}$ constraint,
together with a contribution from supermassive black hole binaries.
While the dilaton and dark glueballs can be produced after the phase transition,
they immediately decay into dark radiation,
which can help ameliorate the Hubble tension
and be tested by the future CMB-S4 experiment.
Alternatively, for a dark conformal sector decaying into the visible sector after the phase transition,
the $\Delta N_{\rm eff}$ constraint is not applied and
the phase transition can solely explain the NANOGrav signal.

\end{abstract}

\maketitle

\renewcommand{\thefootnote}{\arabic{footnote}}
\setcounter{footnote}{0}

%%%%%%%%%%%%%%%%%%%%%%%%%%%%%%%%%%%%%%%%%%%%%%%%%%%
\section{Introduction}\label{introduction}
%%%%%%%%%%%%%%%%%%%%%%%%%%%%%%%%%%%%%%%%%%%%%%%%%%%

The discovery of gravitational waves (GWs) with very low frequencies of $\mathcal{O} (10^{-9}) \, \rm Hz$
will have a major impact
on the understanding of our Universe,
and they are targets of pulsar timing array experiments.
The NANOGrav collaboration~\cite{McLaughlin:2013ira,Brazier:2019mmu} has recently found a positive evidence
for the existence of a stochastic GW background
around $f \sim 10^{-8} \, \rm Hz$ in their 15 years of data~\cite{NANOGrav:2023gor}
(which we call the NANOGrav signal in the following).
Furthermore, a marginal evidence for a GW background has been reported
by the European Pulsar Timing Array (EPTA) collaboration, combining with the first data release of the Indian Pulsar Timing Array (InPTA)
\cite{Antoniadis:2023ott, Antoniadis:2023xlr}.
The reported signal is also consistent with the results of the Parkes Pulsar Timing Array (PPTA)~\cite{Reardon:2023gzh} and
Chinese Pulsar Timing Array (CPTA)~\cite{Xu:2023wog} collaborations.
The NANOGrav signal of a low frequency GW background can be interpreted in terms of inspiraling supermassive black hole binaries (SMBHBs)
\cite{Middleton:2020asl,NANOGrav:2020spf} (see ref.~\cite{Ellis:2023dgf} for possible environmental effects on the SMBHB evolution) and/or cosmological sources~\cite{Madge:2023cak} such as 
cosmic strings
\cite{Blasi:2020mfx,Ellis:2020ena} (their implications for grand unification have been studied in ref.~\cite{Chigusa:2020rks}),
cosmological phase transitions
\cite{Nakai:2020oit,Addazi:2020zcj,Ratzinger:2020koh,Neronov:2020qrl,Vagnozzi:2023lwo,Barman:2020jrf,Li:2021qer,Brandenburg:2021tmp,Borah:2021ocu,DiBari:2021dri,Lewicki:2021xku, Ashoorioon:2022raz,Freese:2022qrl,Freese:2023fcr,Bringmann:2023opz,Han:2023olf} (for a recent review see~\cite{Athron:2023xlk}) and so on
(see their reference lists for the original papers pointing out the production of a low frequency GW background).
Interestingly, in ref.~\cite{NANOGrav:2023hvm}, the NANOGrav collaboration claims that many cosmological models,
including the cosmic inflation and first-order phase transitions,
provide a better fit of their data than baseline SMBHBs.
Then, in the present paper, we further explore the possibility to explain the reported GW signal 
by one of the promising cosmological sources, a first-order phase transition that takes place
in the early Universe, with a concrete model of particle physics.

The authors of ref.~\cite{Nakai:2020oit} have discussed the interpretation of the NANOGrav signal
with the 12.5-year data set in terms of
a GW background generated by a first-order phase transition in a dark sector
thermally decoupled from the visible Standard Model (SM) sector.
The temperatures of both sectors are then different, which makes it possible to realize
the viable ratio of the energy densities of the two sectors that leads to
the NANOGrav signal without contradicting the $\Delta N_{\rm eff}$ constraint.
They considered that the remnant of the dark sector after the phase transition is given by dark radiation, 
and pointed out that the confirmation of the NANOGrav signal may indicate
the existence of such a dark radiation component in the Universe,
which can help mitigate the Hubble tension
\cite{Bernal:2016gxb,Planck:2018vyg,Blinov:2020hmc,Vagnozzi:2019ezj}
and be probed by the future CMB-S4 experiment
\cite{CMB-S4:2016ple}.
While their discussion has not specified the concrete shape of the dark sector,
it is not trivial to give a viable first-order phase transition that is able to
generate a sizable GW background indicated by the reported NANOGrav signal.
In fact, for example, confinement-deconfinement and chiral phase transitions in QCD-like theories
generally lead to a GW background whose amplitude does not reach the strength of the signal of NANOGrav,
according to the effective theory argument
(see $e.g.$ refs.~\cite{Aoki:2017aws,Kang:2021epo,Reichert:2021cvs,Fujikura:2023fbi,Morgante:2022zvc}; also see ref.~\cite{Sagunski:2023ynd} for possible enhancement due to QCD-driven strong supercooling). For a discussion on primordial black hole production in confining phase transitions, see refs.~\cite{Lu:2022yuc, Dvali:2021byy}.

The purpose of the present paper is to provide a concrete model to realize the scenario discussed in ref.~\cite{Nakai:2020oit}
and re-analyze the interpretation of the NANOGrav signal in terms of a first-order phase transition by using
the latest result with the 15-year data set~\cite{NANOGrav:2023gor}.
We consider a confining first-order phase transition
of a nearly-conformal dark sector initially thermally decoupled from the visible sector.
Such a nearly-conformal sector generically leads to a strong first-order phase transition~\cite{Witten:1980ez,Buchmuller:1990ds}, which is a suitable candidate of the GW background
detected by NANOGrav.
According to the AdS/CFT correspondence
\cite{Maldacena:1997re,Gubser:1998bc,Witten:1998qj} (see also refs.~\cite{Arkani-Hamed:2000ijo,Rattazzi:2000hs}),
such a conformal sector can be explicitly described by 
its 5D holographic picture of the Randall-Sundrum (RS) model
\cite{Randall:1999ee} whose extra dimension is bounded by two 3-branes called UV and IR branes.
Unlike the original RS model, our visible sector is localized on the UV brane,
and the typical energy scale of the IR brane is hierarchically smaller than the weak scale. 
It has been pointed out in ref.~\cite{Creminelli:2001th}
that the Universe based on the RS model is described by
the AdS-Schwarzschild (AdS-S) spacetime with the IR brane replaced by an event horizon at high temperatures. 
As the temperature decreases,
a phase transition from the AdS-S spacetime to the RS spacetime takes place,
which corresponds to the confining phase transition of the conformal sector in the 4D description.
The phase transition is known to be the strongly first-order and analyzed by looking into a radion potential
obtained by some mechanism to stabilize the size of the extra dimension.
In the present paper, we utilize the radion stabilization mechanism proposed in ref.~\cite{Fujikura:2019oyi},
which introduces a 5D pure Yang-Mills field.
This mechanism has the advantage of completing the phase transition,
while it is generally problematic in the most popular Goldberger-Wise mechanism
\cite{Goldberger:1999uk}.
In the AdS/CFT, the existence of the IR brane corresponds to spontaneous breaking of the conformal symmetry.
The associated Nambu-Goldstone mode $\phi$ is called dilaton and identified as the radion in the 5D picture.
Therefore, a dark pure Yang-Mills field couples to our conformal sector and 
generates a dilaton potential that we use to analyze the phase transition leading to the NANOGrav GW signal.
We will also study the fate of the dilaton and dark glueballs produced after the phase transition.

The rest of the paper is organized as follows.
Section~\ref{model} presents our model of a nearly-conformal sector coupled to a dark pure Yang-Mills field.
We consider a 5D holographic picture that gives a concrete weakly-coupled description of the model
to explicitly show a radion effective potential
that describes a confining first-order phase transition of the conformal sector.
In section~\ref{PT_section}, we discuss the phase transition and present the formulae for its rate.
Section~\ref{GW_section} analyzes a GW background generated by the phase transition.
For a dark sector only gravitationally connected with the visible sector,
we will show that the NANOGrav signal is explained by the phase transition
without contradicting the $\Delta N_{\rm eff}$ constraint,
together with a contribution from supermassive black hole binaries.
The dilaton and dark glueballs produced after the phase transition immediately decay into dark radiation,
which can help ameliorate the Hubble tension and be tested by the future CMB-S4 experiment.
Furthermore, we will discuss the case of a dark conformal sector decaying into the visible sector after the phase transition.
In this scenario, the $\Delta N_{\rm eff}$ constraint is not relevant and
the phase transition can solely explain the reported GW signal.
Section~\ref{conclusion} is devoted to conclusions and discussions.

%%%%%%%%%%%%%%%%%%%%%%%%%%%%%%%%%%%%%%%%%%%%%%%%%%%
\section{The model}\label{model}
%%%%%%%%%%%%%%%%%%%%%%%%%%%%%%%%%%%%%%%%%%%%%%%%%%%

We consider a 4D conformal dark sector initially thermally decoupled from the visible SM sector.%
\footnote{It means we assume that our dark sector is thermally decoupled from the visible sector {\it before}
the conformal phase transition.}
After inflation, both sectors are reheated by the inflaton decay,
and the visible and dark sectors have their own temperatures denoted as $T$ and $T^{(D)}$, respectively. 
For the dark sector, the system at high temperature is described by a thermal CFT.
To trigger the confinement of the CFT,
we introduce a dark $SU(N_H)$ pure Yang-Mills field coupled to the conformal sector.
Due to asymptotic freedom, its effect on the conformal sector is negligible at high energies
while it becomes significant at low energies.
The confinement of the dark pure Yang-Mills field drives spontaneous breaking of the conformal symmetry generating a mass gap
in the dark sector.
As the temperature $T^{(D)}$ decreases,
a confinement-deconfinement phase transition then takes place, which finally leads to a GW background
explaining the NANOGrav signal.
The phase transition is analyzed by looking into a dilaton effective potential $V_{\rm eff} (\phi)$
generated by the dark Yang-Mills field.

To give a concrete weakly-coupled description of our scenario,
we consider a holographic model with a warped extra dimension bounded by two 3-branes~\cite{Randall:1999ee}.
At zero temperature, the 5D warped geometry is given by $\mathbb{R}^4\times S_1/\mathbb{Z}_2$,
and the background metric is
\begin{align}
\label{eq:RS_metric}
    ds^2=e^{-2k y}\eta_{\mu\nu}dx^\mu dx^\nu +dy^2\ .
\end{align}
Here, $x^\mu \, (\mu = 0,1,2,3)$ and $y$ denote the $3+1$ spacetime and extra space coordinates,
and $k$ is the AdS curvature of $\mathcal{O}(M_{\rm Pl})$.
The UV and IR branes are located at orbifold fixed points of $y=0,~y_{\rm IR}$, respectively.
The typical energy scale at the IR brane is exponentially suppressed compared to that of the UV brane
due to the warp factor $e^{-k y_{\rm IR}}$.
We assume that all the SM fields live on the {UV} brane.
According to the AdS/CFT, the AdS$_5$ bulk geometry is dual to a 4D CFT
with the number of colors $N \equiv 4 \pi \left(M_5/k\right)^{3/2}$,
where $M_{5}$ denotes the 5D Planck scale.
That is our dark conformal sector.
$N$ should be large, $N\gtrsim 4.4$, so that higher order terms of the Ricci
scalar are safely neglected~\cite{Agashe:2007zd}.

The dilaton $\phi$ is identified as a radion degree of freedom 
corresponding to the distance between the two branes in the 5D picture.
To generate a radion potential to stabilize the distance, we introduce a 5D $SU(N_H)$ pure Yang-Mills field.
For details of the derivation of the potential, see refs.~\cite{vonHarling:2017yew,Fujikura:2019oyi}.
The 4D radion/dilaton effective action is given by
\begin{equation}
	S_{\rm eff} = \int d^4x \left[  \frac{3 N^2}{4 \pi^2}  ( \partial \phi (x))^2 -  V_{\rm eff}(\phi) \right]  ,
	\label{radion_kin:eq}
\end{equation} 
where
\begin{equation}
\label{radion potential}
    V_{\rm eff}(\phi)=
    \left\{
\begin{array}{ll}
V_0+\frac{\lambda}{4}\phi^4-\frac{b_{\rm YM}}{8}\Lambda^4_{H,0}\left(\frac{\phi}{\phi_{\rm min}}\right)^{4n} & \text{for}~\phi \geq \phi_c  \\[2ex]
V_0+\frac{\lambda}{4}\phi^4-\frac{b_{\rm YM}}{8}\gamma_c^4\phi_c^4 & \text{for}~\phi<\phi_c \ .
\end{array}
\right.
\end{equation}
Here, the field value at the potential minimum $\phi_{\rm min}$
determines the scale of spontaneous breaking of the conformal symmetry.
The constant $V_0$ is tuned to reproduce the observed vanishingly-small cosmological constant,
$\lambda$ is a dimensionless constant originated from the IR brane tension,
$b_{\rm YM}=11N_H/3$ is the beta function coefficient of the dark Yang-Mills field,
$\Lambda_{H,0}$ denotes the confinement scale at the minimum of the potential, $\gamma_c\approx\pi$ and $n<1$.
The behavior of the potential changes at $\phi=\phi_c$ with
\begin{align}
    \phi_c = \phi_{\rm min} \left( \frac{\Lambda_{H,0}}{\gamma_c \phi_{\rm min} }\right)^{1/(1-n)} ,
\end{align}
because the confinement scale of the $SU(N_H)$ gauge field is independent of the value of $\phi$ for $\phi < \phi_c$.
The validity of the potential \eqref{radion potential} requires $\phi_{\rm min} > \phi_c$.

We consider two scenarios for the fate of our dark sector after the conformal phase transition.
The first is to assume all the energies stored in the dark sector go into dark radiation. 
In this scenario, for instance, a massless dark photon localized on the IR brane can be introduced.
Such a dark photon field couples to the radion through the trace of the energy-momentum tensor.
The coupling is only suppressed by the typical mass scale on the IR brane, $M_{\rm IR}\sim k e^{-k y_{\rm IR}}$.
Then, the radion quickly decays into the dark photon
once the dark sector temperature $T^{(D)}$ becomes lower than the radion mass.
Kaluza-Klein (KK) gravitons also couple to fields on the IR brane through the energy-momentum tensor,
and decay just after the phase transition.
The dark Yang-Mills field forms glueballs in the confinement phase.
Assuming that the CP symmetry is broken in the dark sector,
those glueballs also safely decay into dark radiation through the mixing with the radion.
The resulting dark radiation component in the Universe is subject to the $\Delta N_{\rm eff}$ constraint.

Alternatively, we can assume that all the dark sector energies quickly go into the visible sector after the phase transition.
In this scenario, for instance, a scalar degree of freedom existing after the transition
mixes with the SM Higgs due to some portal coupling so that the dark sector decays.
Indeed, a scalar field with mass of around MeV and lifetime of around $10^{-2}-10^{-1}$\,s
is able to be consistent with the Big Bang Nucleosynthesis (BBN) and laboratory experiments according to ref.~\cite{Bringmann:2023opz}.
Another possibility is to introduce a contact term between the 5D $SU(N_H)$ gauge field and the SM Higgs field $H$ on the UV brane, ${G_H G_H}|H|^2$ where $G_H$ denotes the field strength of the  $SU(N_H)$ gauge field. This interaction provides dark glueball couplings to the visible sector, and enables the radion and dark glueballs to decay into SM particles after the phase transition. 
It is still a non-trivial issue to construct a concrete model in which such decays do not ruin the prediction of the BBN and
the consistency with other experiments, which is left for a future study. 
Another subtle point in this scenario is whether the dark sector is decoupled from the visible sector
before the phase transition.
That is, if the decays into the visible sector are too fast,
it indicates the existence of a relatively large coupling between the two sectors,
and the ``dark'' sector may form the thermal bath with the visible sector even before the phase transition.
Nonetheless, we consider this second scenario, as well as the first scenario,
regardless of the naturalness or fine-tuning of model parameters.

%%%%%%%%%%%%%%%%%%%%%%%%%%%%%%%%%%%%%%%%%%%%%%%%%%%
\section{Phase Transition}\label{PT_section}
%%%%%%%%%%%%%%%%%%%%%%%%%%%%%%%%%%%%%%%%%%%%%%%%%%%
%==================================================

At finite temperature, the geometry of the 5D spacetime, whose time direction is Euclidean and compactified on a circle,
admits two different phases, one of which is energetically favorable over the other,
depending on the temperature~\cite{Creminelli:2001th}.
At sufficiently high temperature $T^{(D)}$ compared to the typical mass scale of the radion potential, the AdS-S spacetime in which the IR brane is replaced by an event horizon is thermodynamically favored.
Its metric is given by
\begin{align}
    ds^2=\left(\frac{\rho^2}{L^2}-\frac{\rho_h^4/L^2}{\rho^2}\right)dt^2+\frac{d\rho^2}{\frac{\rho^2}{L^2}-\frac{\rho_h^4/L^2}{\rho^2}}+\frac{\rho^2}{L^2}\sum_i dx^2_i\ .
\end{align}
Here, $L=1/k$ and $\rho\geq \rho_h$ where the event horizon is at $\rho=\rho_h$.
The above metric becomes a solution of the Einstein equation when the time direction has a periodicity of
$\beta=\pi L^2/\rho_h$ where $\beta$ is the inverse of the temperature $T^{(D)}$.
Otherwise, the conical singularity appears.
At low temperature, the geometry is described according to the RS metric~\eqref{eq:RS_metric} with the given time periodicity.
As we will see later, the RS spacetime with a stabilized radion becomes thermodynamically favored at low temperature.
Therefore, a phase transition is expected to take place in the dark sector as it evolves and cools below the scale of the 
IR brane.
In the 4D dual description, this is equivalent to a confinement-deconfinement phase transition in the dark conformal sector with an explicit breaking of the scale invariance.

The phase transition between two phases of the AdS-S and RS solutions is studied by estimating the free energies for both sectors.
For the AdS-S phase (or the thermal CFT phase in the dual 4D picture),
the free energy is comprised of a large $N$ CFT
and parameterized by the temperature.
Since the pure AdS solution interpolates between the AdS-S and RS spacetimes, we are interested in the free energy difference between the AdS-S and pure AdS spacetimes,
which is given by
\cite{Creminelli:2001th}
\begin{equation}
    \Delta F_{\rm AdS-S} (T_h) = \frac{3}{8} \pi^2 N^2 T_{h}^4 -\frac{1}{2} \pi^2 N^2 T_{h}^3 T^{(D)} \ ,
    \label{FAds:eq}
\end{equation}
where $T_{h} \, (\equiv k^2 \rho_h / \pi)$ and $T^{(D)}$ denote the Hawking temperature and the ambient temperature in the dark sector, respectively. 
This has a minimum when $T^{(D)}=T_{h}$, and it increases as the ${T^{(D)}}$ decreases.
This free energy difference should be compared with the free energy difference
between the RS spacetime with a stabilized radion and the AdS spacetime, which is given by
\begin{equation}
    \Delta F_{\rm RS} = V_{\rm eff} (\phi_{\rm min}) - V_{\rm eff} (0) \ .
    \label{FRS:eq}
\end{equation}
Here, we assume that a dark gauge field contribution to the free energy is negligible,
which may be justified for $N>N_H$.

Below a critical temperature, it becomes energetically favorable to nucleate IR brane bubbles.
The critical temperature is defined as the temperature where the free energy of the AdS-S spacetime
is equal to that of the RS spacetime, namely, $\Delta F_{\rm AdS-S}(T_c^{(D)})=\Delta F_{\rm RS}$.
This condition turns out to be
\begin{equation}
  T_c^{(D)} = \left(\frac{8 |\Delta F_{\rm RS} |}{\pi^2 N^2} \right)^{1/4} \ .
  \label{critical_temp:eq}
\end{equation}
The nature of the phase transition differs for different radion stabilization mechanisms.
For instance, the standard Goldberger-Wise mechanism leads to a too shallow potential of the radion
due to the approximate scale invariance,
which is not suitable to complete the phase transition.
Conversely, we can see that our radion potential \eqref{radion potential}
generated by a dark Yang-Mills field correctly completes the phase transition
and even avoids a significant supercooling phase
\cite{Fujikura:2019oyi} because the scale invariance is significantly broken by the confinement.

The tunneling rate of the first-order phase transition per unit time and volume receives a dominant contribution
from the ``bounce'', and has the form $\Gamma \simeq A e^{-S_{\rm E}}$,
where $A$ encodes the effects of thermal and quantum fluctuations
and $S_{\rm E}$ denotes the Euclidean action of the bounce solution~\cite{Coleman_bounce}.
The $O(4)$ symmetric bounce action is relevant when the nucleation temperature is much smaller than the critical temperature, as the $O(3)$ symmetric action receives an explicit $1/T^{(D)}$ enhancement. The $O(4)$ symmetric bubble action can be estimated from the radion kinetic normalization and knowledge of the potential barrier and width. In the thick wall approximation, applicable at temperature much lower than the critical temperature,
it is given by
\cite{Creminelli:2001th, Fujikura:2019oyi}
    \begin{align}
        S_4 &\simeq \frac{9 N^4}{8 \pi^2} \frac{(\phi_t + \sqrt{c} \, T^{(D)})^4}{\Delta F_{\rm RS} \left( {T^{(D)}}/{T_c^{(D)}} \right)^4 -\Big[V_{\rm eff} (\phi_t)-V_{\rm eff}(0)\Big]} \ .
        \label{S:eq}
    \end{align}
Here, the tunneling point $\phi_t$ is obtained by minimizing the bounce action, and $c$ represents some $O(1)$ coefficient for the normalization of the kinetic term for the Hawking temperature.
We have checked that the $S_4$ action computed from Eq.~\eqref{S:eq} gives the same order of magnitude as
that obtained by numerically solving the bounce equation in our interesting parameter space.
The nucleation of bubbles proceeds as soon as the transition rate competes with the Hubble rate
$\Gamma  \simeq H^4 (T_*) \equiv H_*^4$
where the corresponding temperature for the visible SM sector during the phase transition is denoted as $T_*$.
As the visible sector energy density dominates the total energy density of the Universe,
the Hubble rate at the time of the phase transition  is primarily determined by the visible sector, namely
    \begin{equation}
        H^2_*= \frac{1}{3 M_{\rm Pl}^2} \left[\rho_{\rm rad} (T_*) + \rho_{\rm DR} (T_{*i}^{(D)}) \right] \simeq \frac{\rho_{\rm rad} (T_*)}{3 M_{\rm Pl}^2} \ ,
        \label{Hubble:eq}
    \end{equation}
where $T_{*i}^{(D)}$ denotes the temperature of the dark sector just before the phase transition,
and $\rho_{\rm rad}$, $\rho_{\rm DR}$ are the visible and dark sector radiation energy densities, respectively.

%==================================================

%%%%%%%%%%%%%%%%%%%%%%%%%%%%%%%%%%%%%%%%%%%%%%%%%%%
\section{Gravitational wave generation}\label{GW_section}
%%%%%%%%%%%%%%%%%%%%%%%%%%%%%%%%%%%%%%%%%%%%%%%%%%%
%==================================================

In order to analyze the GW spectrum produced by the AdS-S to RS phase transition,
two key quantities need to be evaluated, namely the inverse duration of the phase transition, $\beta \equiv - {dS_{\rm E}}/{dt}|_{t=t_*}$ where $t_*$ denotes the cosmic time when GWs are produced,
and the vacuum energy density of the dark sector released to the total radiation bath,
\begin{equation}
    \alpha' \equiv \frac{\rho_{\rm vac}}{\rho_{\rm rad} (T_*) + \rho_{\rm DR} (T_{*i}^{(D)})} \simeq \frac{V(\phi_t)-V(\phi_{\rm min})}{\pi^2 g_*(T_*) \, T_*^4/30} \ .
    \label{alphap:eq}
\end{equation}
Here, $g_*$ ($g_*^{(D)}$) denotes the effective number of relativistic degrees of freedom for the energy of the visible (dark) sector.

%=============================================================
%
 \begin{figure*}[!ht]
    \subfloat[\label{secluded:fig}]{%
      \includegraphics[width=0.46\textwidth]{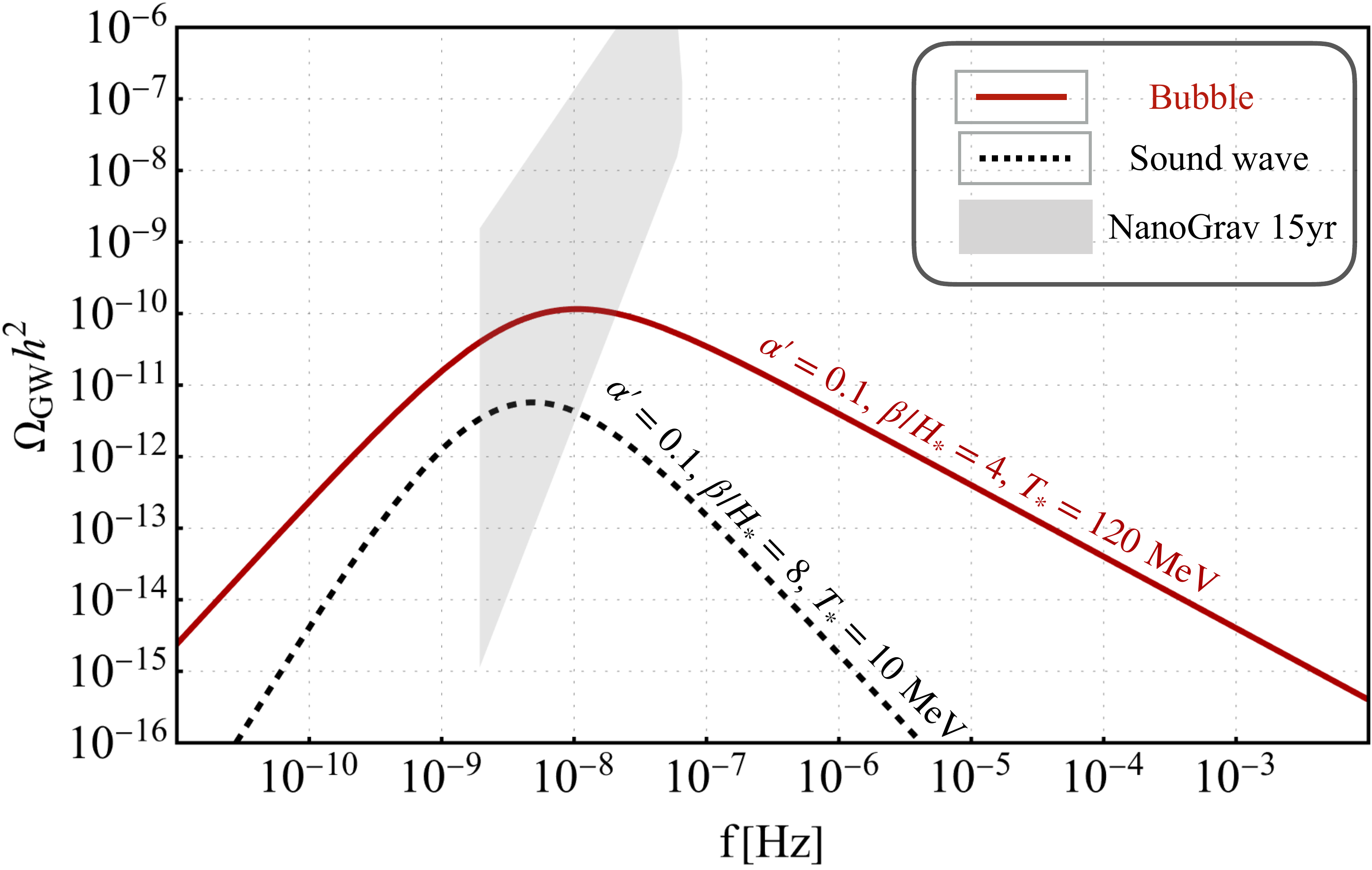}
    }
    \hfill
    \subfloat[\label{decaying:fig}]{%
      \includegraphics[width=0.46\textwidth]{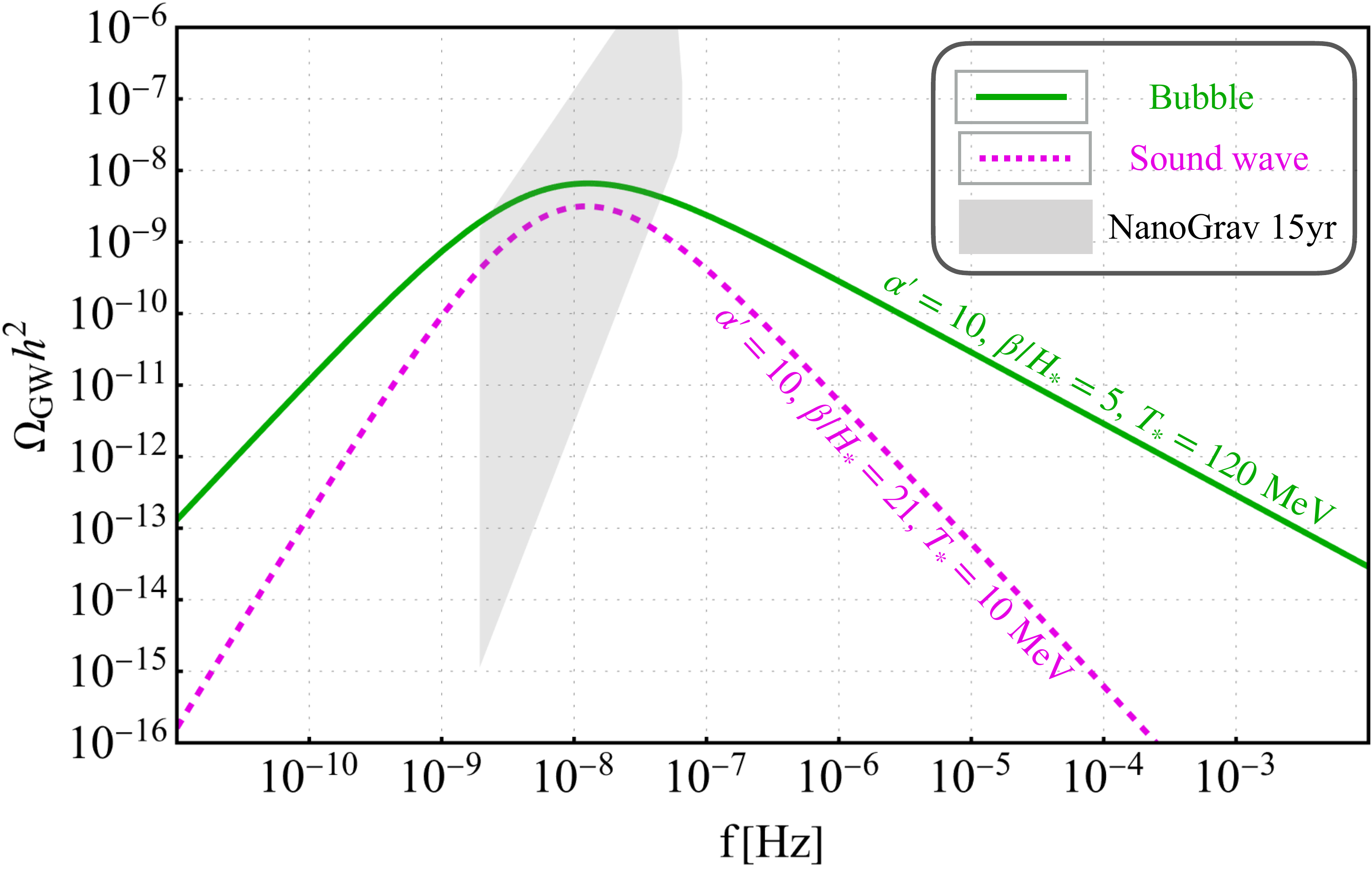}
    }
    \caption{
 A GW background produced by the AdS-S to RS phase transition. The sub-figure~\ref{secluded:fig} (\ref{decaying:fig}) depicts the case where the dark sector is secluded (decays to the SM). In each sub-figure the solid (dashed) line represents the bubble collision (sound wave) only case. For the bubble collision only case in subfig.~\ref{secluded:fig}, we take $N=21$, $N_{H}=10$, $n\equiv 3 N/11 N_{H} = 0.57$, $\lambda=1$, $\gamma_c=\pi$, $T_*=120$ MeV, and $\phi_{\rm min}=0.14$ GeV. For the sound wave only case, we choose $N=21$, $N_{H}=12$, $n\equiv 3 N/11 N_{H} = 0.48$, $T_*=10$ MeV, and $\phi_{\rm min}=10.7$ MeV while the other parameters are identical. Similarly, for subfig.~\ref{decaying:fig}, the bubble only case parameters are $N=21$, $N_{H}=11$, $n\equiv 3 N/11 N_{H} = 0.52$, $T_*=120$ MeV, and $\phi_{\rm min}=0.42$ GeV, while in the sound wave only case $N=21$, $N_{H}=10$, $n\equiv 3 N/11 N_{H} = 0.57$, $T_*=10$ MeV, and $\phi_{\rm min}=37$ MeV, the other parameters being the same. The gray shaded region refers to the NANOGrav 15-yr signal region~\cite{NANOGrav:2023gor}. }
    \label{gw:figs}
  \end{figure*}

Essentially, main contributions to GW signals can be divided into two pieces: bubble collisions $\Omega_{\rm coll}$, and the sound wave of the plasma $\Omega_{\rm sound}$.
There exists the third contribution from the turbulence of the plasma,
while a numerical calculation~\cite{Caprini:2015zlo} shows that it is subdominant compared to that of the sound wave.
Hence, we omit the contribution from the turbulence.
The contribution from bubble collisions is given by~\cite{Jinno:2016vai} (see also refs.~\cite{Kosowsky:1991ua,Kosowsky:1992rz,Kosowsky:1992vn,Kamionkowski:1993fg,Caprini:2007xq,Huber:2008hg,NANOGrav:2021flc})
\begin{align}
\nonumber
&\Omega_{\rm coll} h^2 \simeq 5.5 \times 10^{-7} \left(\dfrac{20}{g_*(T_*)}\right)^{\frac{1}{3}} \left(\frac{H_*}{\beta} \right)^2   \\&\times \left( \frac{\alpha'}{1+\alpha'}\right)^2\mathcal{S}(f/f_{\rm coll})\ ,\nonumber
\\[1ex]
&\mathcal{S}(x)=\dfrac{1}{\mathcal{N}}\dfrac{(a+b)^c}{\big[b x^{-a/c} + a x^{b/c}\big]^c},\label{bubble:eq}\\[1ex]
&\mathcal{N} = \left(\dfrac{b}{a}\right)^{a/n}\left(\dfrac{nc}{b}\right)^c \dfrac{\Gamma(a/n)\Gamma(b/n)}{n\Gamma(c)}, \quad n=\dfrac{a+b}{c}.\nonumber
\end{align}
Here $\Gamma$ is the Euler Gamma function, $(a, b, c)$ denote bubble spectral shape parameters.
For bubble collisions, we take these parameters to be the corresponding to maximum posterior values obtained by the recent NANOGrav analysis~\cite{NANOGrav:2023hvm}, namely $(a,b,c)=(2.01, 1, 2.93)$ for the secluded dark sector scenario
(where a contribution from SMBHBs is assumed)
and $(1.97, 1, 3 )$ for the decaying dark sector scenario.
The sound wave contribution is given by~\cite{Hindmarsh:2013xza,Hindmarsh:2015qta,Hindmarsh:2017gnf}
\begin{align}
\Omega_{\rm sound} & h^2 \simeq 1.4\times 10^{-6}v_w\left(\dfrac{20}{g_*(T_*)}\right)^{\frac{1}{3}}\left(\frac{H_*}{\beta} \right) \\ \nonumber &  \times \left( \frac{\kappa_{\rm sw} \alpha'}{1+\alpha'}\right)^2\left(1-\frac{1}{\sqrt{1+2 t_{\rm sw} H_*}} \right)\mathcal{S}(f/f_{\rm sw})  \ .
\label{sw:eq}
\end{align}
For the sound wave contribution, we also take maximum posterior values, $(a,b,c)= (3, 2, 5)$~\cite{NANOGrav:2023hvm}.
The peak frequencies $f_{\rm coll,sw}$ are given by 
\begin{align}
&f_{\rm coll,sw}\simeq 126\,{\rm nHz}\,
    \frac{1}{v_w}\left(\dfrac{g_*(T_*)}{20}\right)^{\frac{1}{6}}\left(\dfrac{T_*}{1\,{\rm GeV}}\right)\dfrac{\beta}{H_*}f^{*}_{\rm coll,sw}\,,\\
    &f^*_{\rm coll}\simeq 0.2  , \quad f^*_{\rm sw}\simeq 0.54 ,  \\[1ex]
  & t_{\rm sw}  \sim (8 \pi)^{1/3} \frac{v_w}{\beta \sqrt{3\kappa_{\rm sw} \alpha'/4(1+\alpha') }} \ .
  \label{GW_amp:eqs}
\end{align} 
We have included the suppression factor of the short-period of the sound wave~\cite{Ellis:2018mja,Ellis:2019oqb,Ellis:2020awk,Guo:2020grp}.
In these expressions, $v_w$ and $\kappa_{\rm sw}$ are the bubble wall velocity and an efficiency factor
encoding the fractions of the latent energy released into the bulk motion of the fluid and into the bubble shell, respectively.
For the efficiency factor $\kappa_{\rm sw}$, we use the following expression \cite{Espinosa:2010hh},
\begin{align}
\kappa_{\rm sw} = \dfrac{\alpha'}{0.73+0.083\sqrt{\alpha'}+\alpha'}.
\end{align}
This expression is suitable for a large bubble wall velocity, $v_w\to 1$.
The total amount of a GW background is given by the sum of two contributions.\footnote{{We note that in the regime $\beta/H \lesssim 10$, numerical simulations may overestimate the strength of the GW signal~\cite{Bringmann:2023opz},
while a more precise analysis is beyond the scope of the present paper.}}

If the interaction between nucleated bubbles and the thermal plasma is strong, most of the kinetic energy of the accelerating bubble wall is believed to be injected into the
thermal bath.
In this case, the sound wave of
the plasma becomes the main source of GWs.
In the other case where the bubble wall does not receive a strong friction from the dark thermal plasma,
its bulk energy would be negligible.
Consequently, bubble collisions become the most dominant source of GWs.
This may take place when the phase transition is strongly supercooled so that the thermal bath is significantly diluted.
In the intermediate case, two sources can be comparable.
In our analysis, we consider both cases, where the most dominant contribution comes from the bubble collisions
or the sound wave of the thermal plasma as in ref.~\cite{NANOGrav:2023hvm}.
Moreover, to precisely estimate the amplitude of GWs, one needs to evaluate the bubble wall velocity $v_w$.
However, the estimation of $v_w$ significantly depends on the interaction of the accelerating wall with the thermal bath, and is still an open theoretical question
(see, however, refs.~\cite{Bodeker:2009qy,Bodeker:2017cim,Dorsch:2018pat,BarrosoMancha:2020fay,Laurent:2020gpg,Azatov:2020ufh,Wang:2020zlf,Gouttenoire:2021kjv} for recent developments).
When bubble collisions are the dominant source of GWs, the plasma effect on the bubble wall is expected to be subdominant, and hence, one may naturally set $v_w=1$.
Following the original NANOGrav analysis~\cite{NANOGrav:2023hvm}, we also use optimistic bubble wall velocity $v_w=1$ in the case where the sound wave contribution is dominant.

In the secluded dark sector scenario, as most of the vacuum energy is injected into dark radiation after the phase transition, $\alpha'$ also determines the fraction of the dark radiation energy density to the visible radiation energy density just after the phase transition.
Such a dark radiation component acts as extra relativistic neutrino species during the recombination epoch, 
\begin{equation}
    \rho_{\rm DR,0} \equiv \frac{7}{8} \Delta N_{\rm eff} \left(\frac{4}{11}\right)^{4/3} \rho_{\gamma,0} \ ,
    \label{Neff:eq}
\end{equation}
where $\rho_{\gamma}$ is the photon energy density and the subscript `$0$' denotes the value at the recombination era.
By using entropy conservation, we can relate $\alpha'$ to $\Delta N_{\rm eff}$ as
\begin{equation}
    \alpha' \simeq 0.07 \left( \frac{\Delta N_{\rm eff}}{0.5}\right) \left( \frac{g_{*0}}{g_{*}}\right) \left( \frac{g_{*s}}{g_{*s0}}\right)^{4/3} \left( \frac{g_{*}^{(D)}}{g_{*0}^{(D)}}\right) \left( \frac{g_{*s0}^{(D)}}{g_{*s}^{(D)}}\right)^{4/3} \!\!\!\!\! ,
    \label{alphap_Neff:eq}
\end{equation}
with $g_{*s}$ ($g_{*s}^{(D)}$) as the effective number of relativistic degrees of freedom for the entropy density of the visible (dark) sector.
{The current joint constraint from {\em Planck} 2018 CMB + lensing + BAO with low-redshift measurement of the Hubble constant~\cite{Planck:2018vyg} gives $N_{\rm eff} = 3.27 \pm 0.15$ (at 68\% C.L.), while the SM theoretical prediction is $N_{\rm eff, SM} \simeq 3.045$~\cite{deSalas:2016ztq}.
The constraint can be subject to systematics, for example, in the CMB polarization data~\cite{Bernal:2016gxb},
and here we take the conservative limit $\Delta  N_{\rm eff} < 0.7$.}

Figure~\ref{gw:figs} shows a GW background produced by the AdS-S to RS phase transition
both for a secluded dark sector (\ref{secluded:fig}) and for a decaying dark sector (\ref{decaying:fig}).
In subfig.~\ref{secluded:fig}, the bubble collision only case is characterized by the GW parameters, $\alpha'=0.1$, $\beta/H_* = 4$ with $T_* = 0.12$ GeV, while the sound wave case corresponds to $\alpha'=0.1$, $\beta/H_* = 8$ with $T_* = 10$ MeV. Both of these give $\Delta N_{\rm eff} \simeq 0.6$, which can ameliorate the Hubble tension, and is still consistent with the phase transition + SMBHB interpretation of the NANOGrav signal~\cite{NANOGrav:2023gor}. The microscopic parameters that result into these are given in the caption of fig.~\ref{gw:figs}. On the other hand, a decaying dark sector is not restricted from the $\Delta N_{\rm eff}$ bound, and hence can result into a larger $\alpha'$, and consequently, a stronger amplitude. This is reflected in subfig.~\ref{decaying:fig}, where the bubble collision only case is endowed with $\alpha'=10$, $\beta/H_*=5$ with $T_* =0.12$ GeV, whereas the sound wave case has $\alpha'=10$, $\beta/H_*=21$ with $T_* =10$ MeV.

%%%%%%%%%%%%%%%%%%%%%%%%%%%%%%%%%%%%%%%%%%%%%%%%%%%
\section{Discussions}\label{conclusion}
%%%%%%%%%%%%%%%%%%%%%%%%%%%%%%%%%%%%%%%%%%%%%%%%%%%

We have considered the interpretation of the reported NANOGrav signal of a GW background
in terms of a confining first-order phase transition of a nearly-conformal dark sector
initially thermally decoupled from the visible sector.
To give a concrete weakly-coupled description of this scenario,
we discussed a holographic model with a warped extra dimension
where our visible SM sector is localized on the UV brane.
Then, the phase transition is described by a radion potential
generated by a new 5D dark Yang-Mills field.
After the phase transition, we considered two scenarios of a secluded dark sector
and a dark sector decaying into the visible SM sector.
For both scenarios, we found a viable parameter space where the phase transition can produce a GW background explaining the NANOGrav signal.
Furthermore, in the secluded dark sector scenario, the resulting dark radiation component can help ameliorate the Hubble tension
and be tested by the future CMB-S4 experiment.

To explain the NANOGrav signal, we required that the vacuum energy density of the nearly-conformal dark sector is comparable to the energy density of the visible sector at the time of the phase transition. This poses a coincidence problem in cosmology.
Another infamous coincidence problem is the tuning of the cosmological constant (CC).
A possible answer to the CC problem is provided by anthropic considerations~\cite{Weinberg:2000yb,Vilenkin:2001bs,Garriga:1999hu,Garriga:2000cv},
while it is unclear if similar considerations might be applicable to the present coincidence.
We only raise this problem here and leave its solution to a future exploration.

As pointed out in ref.~\cite{Lee:2021wau},
we can extend our model by introducing another 3-brane into the 5D spacetime
to address the electroweak naturalness problem at the same time.
The Goldberger-Wise stabilization of such a three 3-brane setup has been discussed in ref.~\cite{Lee:2021wau},
and ref.~\cite{Girmohanta:2023sjv} studied cosmology in the case where the scale of the IR brane
is given by the weak scale
(for recent applications of the multi-brane setup,
see $e.g.$ refs.~\cite{Agashe:2016rle,Lee:2021slp,Fuentes-Martin:2022xnb,Girmohanta:2022giy}).
In this extended model, the visible SM sector is localized on the intermediate brane.
Since the typical scale of the intermediate brane is given by the electroweak scale,
it is a non-trivial requirement to be checked that
the visible and dark sectors are thermally decoupled before the phase transition.

\section*{Acknowledgments}
This work is supported by Natural Science Foundation of China No.~12150610465 (YN), JSPS KAKENHI Grant Numbers JP22J00537 (MS), JP22J00345 (KF).

\bibliography{ref}

\end{document}